\documentclass[twocolumn,showpacs,prl,superscriptaddress]{revtex4-1}

\usepackage{graphicx}
\RequirePackage{lineno}
\usepackage{color}
\usepackage{cancel}
\usepackage{ulem}

\newcommand{\snn}{\ensuremath{\sqrt{s_{_{\rm NN}}}}}
\newcommand{\vthree}{\ensuremath{v_3^2\{2\}}}

\newcommand{\npvthree}{\ensuremath{N_{\mathrm{part}}v_3^2\{2\}}}
\newcommand{\npart}{\ensuremath{N_{\mathrm{part}}}}

\begin{document}


\title{Beam Energy Dependence of the Third Harmonic of Azimuthal Correlations in Au+Au Collisions at RHIC}

\affiliation{AGH University of Science and Technology, FPACS, Cracow 30-059, Poland}
\affiliation{Argonne National Laboratory, Argonne, Illinois 60439}
\affiliation{Brookhaven National Laboratory, Upton, New York 11973}
\affiliation{University of California, Berkeley, California 94720}
\affiliation{University of California, Davis, California 95616}
\affiliation{University of California, Los Angeles, California 90095}
\affiliation{Central China Normal University, Wuhan, Hubei 430079}
\affiliation{University of Illinois at Chicago, Chicago, Illinois 60607}
\affiliation{Creighton University, Omaha, Nebraska 68178}
\affiliation{Czech Technical University in Prague, FNSPE, Prague, 115 19, Czech Republic}
\affiliation{Nuclear Physics Institute AS CR, 250 68 Prague, Czech Republic}
\affiliation{Frankfurt Institute for Advanced Studies FIAS, Frankfurt 60438, Germany}
\affiliation{Institute of Physics, Bhubaneswar 751005, India}
\affiliation{Indian Institute of Technology, Mumbai 400076, India}
\affiliation{Indiana University, Bloomington, Indiana 47408}
\affiliation{Alikhanov Institute for Theoretical and Experimental Physics, Moscow 117218, Russia}
\affiliation{University of Jammu, Jammu 180001, India}
\affiliation{Joint Institute for Nuclear Research, Dubna, 141 980, Russia}
\affiliation{Kent State University, Kent, Ohio 44242}
\affiliation{University of Kentucky, Lexington, Kentucky, 40506-0055}
\affiliation{Korea Institute of Science and Technology Information, Daejeon 305-701, Korea}
\affiliation{Institute of Modern Physics, Chinese Academy of Sciences, Lanzhou, Gansu 730000}
\affiliation{Lawrence Berkeley National Laboratory, Berkeley, California 94720}
\affiliation{Max-Planck-Institut fur Physik, Munich 80805, Germany}
\affiliation{Michigan State University, East Lansing, Michigan 48824}
\affiliation{National Research Nuclear Univeristy MEPhI, Moscow 115409, Russia}
\affiliation{National Institute of Science Education and Research, Bhubaneswar 751005, India}
\affiliation{National Cheng Kung University, Tainan 70101 }
\affiliation{Ohio State University, Columbus, Ohio 43210}
\affiliation{Institute of Nuclear Physics PAN, Cracow 31-342, Poland}
\affiliation{Panjab University, Chandigarh 160014, India}
\affiliation{Pennsylvania State University, University Park, Pennsylvania 16802}
\affiliation{Institute of High Energy Physics, Protvino 142281, Russia}
\affiliation{Purdue University, West Lafayette, Indiana 47907}
\affiliation{Pusan National University, Pusan 46241, Korea}
\affiliation{University of Rajasthan, Jaipur 302004, India}
\affiliation{Rice University, Houston, Texas 77251}
\affiliation{University of Science and Technology of China, Hefei, Anhui 230026}
\affiliation{Shandong University, Jinan, Shandong 250100}
\affiliation{Shanghai Institute of Applied Physics, Chinese Academy of Sciences, Shanghai 201800}
\affiliation{State University Of New York, Stony Brook, NY 11794}
\affiliation{Temple University, Philadelphia, Pennsylvania 19122}
\affiliation{Texas A\&M University, College Station, Texas 77843}
\affiliation{University of Texas, Austin, Texas 78712}
\affiliation{University of Houston, Houston, Texas 77204}
\affiliation{Tsinghua University, Beijing 100084}
\affiliation{United States Naval Academy, Annapolis, Maryland, 21402}
\affiliation{Valparaiso University, Valparaiso, Indiana 46383}
\affiliation{Variable Energy Cyclotron Centre, Kolkata 700064, India}
\affiliation{Warsaw University of Technology, Warsaw 00-661, Poland}
\affiliation{Wayne State University, Detroit, Michigan 48201}
\affiliation{World Laboratory for Cosmology and Particle Physics (WLCAPP), Cairo 11571, Egypt}
\affiliation{Yale University, New Haven, Connecticut 06520}

\author{L.~Adamczyk}\affiliation{AGH University of Science and Technology, FPACS, Cracow 30-059, Poland}
\author{J.~K.~Adkins}\affiliation{University of Kentucky, Lexington, Kentucky, 40506-0055}
\author{G.~Agakishiev}\affiliation{Joint Institute for Nuclear Research, Dubna, 141 980, Russia}
\author{M.~M.~Aggarwal}\affiliation{Panjab University, Chandigarh 160014, India}
\author{Z.~Ahammed}\affiliation{Variable Energy Cyclotron Centre, Kolkata 700064, India}
\author{I.~Alekseev}\affiliation{Alikhanov Institute for Theoretical and Experimental Physics, Moscow 117218, Russia}
\author{A.~Aparin}\affiliation{Joint Institute for Nuclear Research, Dubna, 141 980, Russia}
\author{D.~Arkhipkin}\affiliation{Brookhaven National Laboratory, Upton, New York 11973}
\author{E.~C.~Aschenauer}\affiliation{Brookhaven National Laboratory, Upton, New York 11973}
\author{A.~Attri}\affiliation{Panjab University, Chandigarh 160014, India}
% ...

\collaboration{STAR Collaboration}\noaffiliation


\begin{abstract}
We present results from a harmonic decomposition of two-particle
azimuthal correlations measured with the STAR detector in Au+Au
collisions for energies ranging from {\snn}$=7.7$ GeV to 200 GeV. The
third harmonic $v_3^2\{2\}=\langle \cos3(\phi_1-\phi_2)\rangle$, where
$\phi_1-\phi_2$ is the angular difference in azimuth, is studied as a
function of the pseudorapidity difference between particle pairs
$\Delta\eta = \eta_1-\eta_2$. Non-zero {\vthree} is directly related
to the previously observed large-$\Delta\eta$ narrow-$\Delta\phi$
ridge correlations and has been shown in models to be sensitive to the
existence of a low viscosity Quark Gluon Plasma (QGP) phase. For
sufficiently central collisions, $v_3^2\{2\}$ persist down to an
energy of 7.7 GeV suggesting that QGP may be created even in these low
energy collisions. In peripheral collisions at these low energies
however, $v_3^2\{2\}$ is consistent with zero. When scaled by
pseudorapidity density of charged particle multiplicity per
participating nucleon pair, $v_3^2\{2\}$ for central collisions shows
a minimum near {\snn}$=20$ GeV.

\end{abstract}


\maketitle


Researchers collide heavy nuclei at ultra-relativistic energies to
create nuclear matter hot enough to form a Quark Gluon Plasma
(QGP)~\cite{Collins:1974ky,Chin:1978gj,Kapusta:1979fh,Anishetty:1980zp};
QGP permeated the entire universe in the first few microseconds after
the Big Bang. Lattice QCD calculations show that the transition
between hadronic matter and a QGP at zero baryon chemical potential is
a smooth cross-over~\cite{Aoki:2006we}. Data from the Relativistic
Heavy Ion Collider (RHIC) at Brookhaven National Laboratory and at the
Large Hadron Collider (LHC) at CERN have been argued to show that the
matter created in these collisions is a nearly perfect fluid with a
viscosity-to-entropy density ratio smaller than any other fluid known
in
nature~\cite{wp,perfect,Gale:2012rq,Chatrchyan:2013kba,Abelev:2014pua}.
At the higher collision energies, baryon number is not as easily
transported from beam rapidity to mid-rapidity leaving the matter at
mid-rapidity nearly net baryon free~\cite{Adler:2001bp}. As {\snn} is
decreased however, more baryon number can be transported to
mid-rapidity creating a system with a larger net baryon density and
larger baryon chemical potential
($\mu_B$)~\cite{Appelshauser:1998yb,Adams:2003xp,Abelev:2008ab}. Collisions
with higher $\mu_B$ values probe a region of the temperature-$\mu_B$
phase diagram, where the transition between QGP and hadrons may change
from a smooth cross-over to a first-order phase
transition~\cite{criticalpoint}, thus defining a possible critical
point. In addition to having a larger $\mu_B$, collisions at lower
{\snn} will also start with lower initial temperatures. For this
reason, the system will spend relatively more time in the
transition region until, at low enough {\snn}, it will presumably fail
to create a QGP. It is not currently known at what $\mu_B$ the
transition might become first-order or at what {\snn} the collision
region will become too cold to create a QGP. In this letter, we report
on measurements of particle correlations that are expected to be
sensitive to whether a low viscosity QGP phase has been created.

Correlations between particles emitted from heavy-ion collisions are
particularly rich in information about the dynamics of the
collision. It has been found that pairs of particles are
preferentially emitted with small relative azimuthal angles
($\Delta\phi = \phi_1-\phi_2 \sim 0$)~\cite{ridgedata}. Surprisingly,
this preference persists even when the particles are separated by
large pseudo-rapidity ($\eta$) gaps ($\Delta\eta$$\gg$$0$). These
long-range correlations, known as the ridge, have been traced to the
conversion of density anisotropies in the initial overlap of the two
nuclei into momentum space correlations through subsequent
interactions in the
expansion~\cite{radflow,Mishra:2007tw,Sorensen:2008dm,Takahashi:2009na,AR}.
Hydrodynamic models have been shown to require a low viscosity
  plasma phase early in the evolution to propagate the geometry
fluctuations through pressure gradients into correlations between
particles produced at freeze-out~\cite{perfect,Gale:2012rq}. Reduction
in the pressure, as expected during a mixed phase for example, should
lead to a reduction in the observed
correlations~\cite{Sorge:1998mk,Voloshin:1999gs,Stoecker:2004qu,Adamczyk:2014ipa}.
The strength of correlations at different length scales can be studied
through the analysis of $v_n^2\{2\}=\langle\cos n(\Delta\phi)\rangle$
as a function of $\Delta\eta$. The second harmonic in this
decomposition is dominated by asymmetries related to the elliptic
shape of the collision overlap region and has been studied for
decades~\cite{v2papers,reviews}. The higher harmonics in this
decomposition received attention more
recently~\cite{ridgedata,earlyv3,Adare:2011tg,Adamczyk:2013waa} after
the importance of the initial density fluctuations was
realized~\cite{radflow,Mishra:2007tw,Sorensen:2008dm,Takahashi:2009na,AR}. The
harmonic $v_3^2\{2\}$ is thought to be particularly sensitive to the
presence of a QGP phase: Hybrid model calculations show that while the
large elliptic shape of the overlap region can develop into
$v_2^2\{2\}$ throughout the evolution, including the hadronic phase,
the development of $v_3^2\{2\}$ relies more strongly on the presence
of a low viscosity QGP phase early in the
collision~\cite{Auvinen:2013sba,Solanki:2012ne}. This suggests unless
an alternative explanation for {\vthree} is found~\cite{He:2015hfa},
$v_3^2\{2\}$ will be an ideal observable to probe the formation of a
QGP and the pressure gradients in the early plasma phase. In this
letter we present measurements of $v_3^2\{2\}(\Delta\eta)$ as a
function of centrality in Au+Au collisions at {\snn}$= 7.7, 11.5,
14.5, 19.6, 27, 39, 62.4$ and 200 GeV by the STAR detector at RHIC. We
also compare these measurements to similar measurements from 2.76 TeV
Pb+Pb collisions at the LHC~\cite{earlyv3}.

The charged particles used in this analysis are detected through
ionization energy loss in the STAR Time Projection
Chamber~\cite{STAR}. The transverse momentum $p_T$, $\eta$, and charge
are determined from the trajectory of the track in the solenoidal
magnetic field of the detector. With the 0.5 Tesla magnetic field used
during data taking, particles can be reliably tracked for $p_T>0.2$
GeV/c. The efficiency for finding particles drops quickly as $p_T$
decreases below this value~\cite{Abelev:2008ab}. Weights $w_{i,j}$ have been
used to correct the correlation functions for the $p_T$-dependent
efficiency and for imperfections in the detector acceptance.  The
quantity analyzed and reported as $v_{n}^{2}\{2\}(\Delta\eta)$ is
\begin{equation}
\langle \cos n(\Delta\phi) \rangle =  \left\langle\left( \frac{\sum_{i,j, i \ne j} w_iw_j\cos n(\phi_i-\phi_j)}{\sum_{i,j, i \ne j}w_iw_j}\right)\right\rangle
\end{equation}
where $\sum_{i,j, i \ne j}$ is a sum over all unique pairs in an
  event and $\langle...\rangle$ represents an average over events with
  each event weighted by the number of pairs in the event.
The weights $w_{i,j}$ are determined from the
inverse of the $\phi$ distributions after they have been averaged over
many events (which for a perfect detector, should be flat) and by the
$p_T$ dependent efficiency. The $w_{i,j}$ depend on the $p_T$, $\eta$,
and charge of the particle, the collision centrality, and the
longitudinal position of the collision vertex. The correction
procedure is verified by checking that the $\phi$ distributions are
flat after the correction and that $\langle\cos n(\phi)\rangle$ and
$\langle\sin n(\phi)\rangle$ are much smaller than the
$\langle\cos(n\Delta\phi)\rangle$~\cite{Bilandzic:2010jr}. With these
corrections applied, the data represent the $v_n^2\{2\}(\Delta\eta)$
that would be seen by a detector with perfect acceptance for particles
with $p_T>0.2$ GeV/c and $|\eta|<1$. Some previous
  results~\cite{Adamczyk:2013waa} on the $\Delta\eta$ dependence of
  {\vthree} use average rather than differential corrections leading
  to small differences in the $\Delta\eta$ dependence between that
  work and this work. The difference is largest in central collisions
  at $1.5<\Delta\eta<2$ where the $v_3^2\{2\}(\Delta\eta)$ reported
  previously is smaller by about 25\%. The difference becomes less
  significant elsewhere. The data have been divided into standard
centrality classes based on the number of charged hadrons observed for
a given event within the pseudo-rapidity region $|\eta|<0.5$. In some
figures, we report the centrality in terms of the number of
participating nucleons ({\npart}) estimated from Monte Carlo Glauber
calculations~\cite{Abelev:2008ab,glauber}.

\begin{figure*}[htb]
\includegraphics[width=\textwidth]{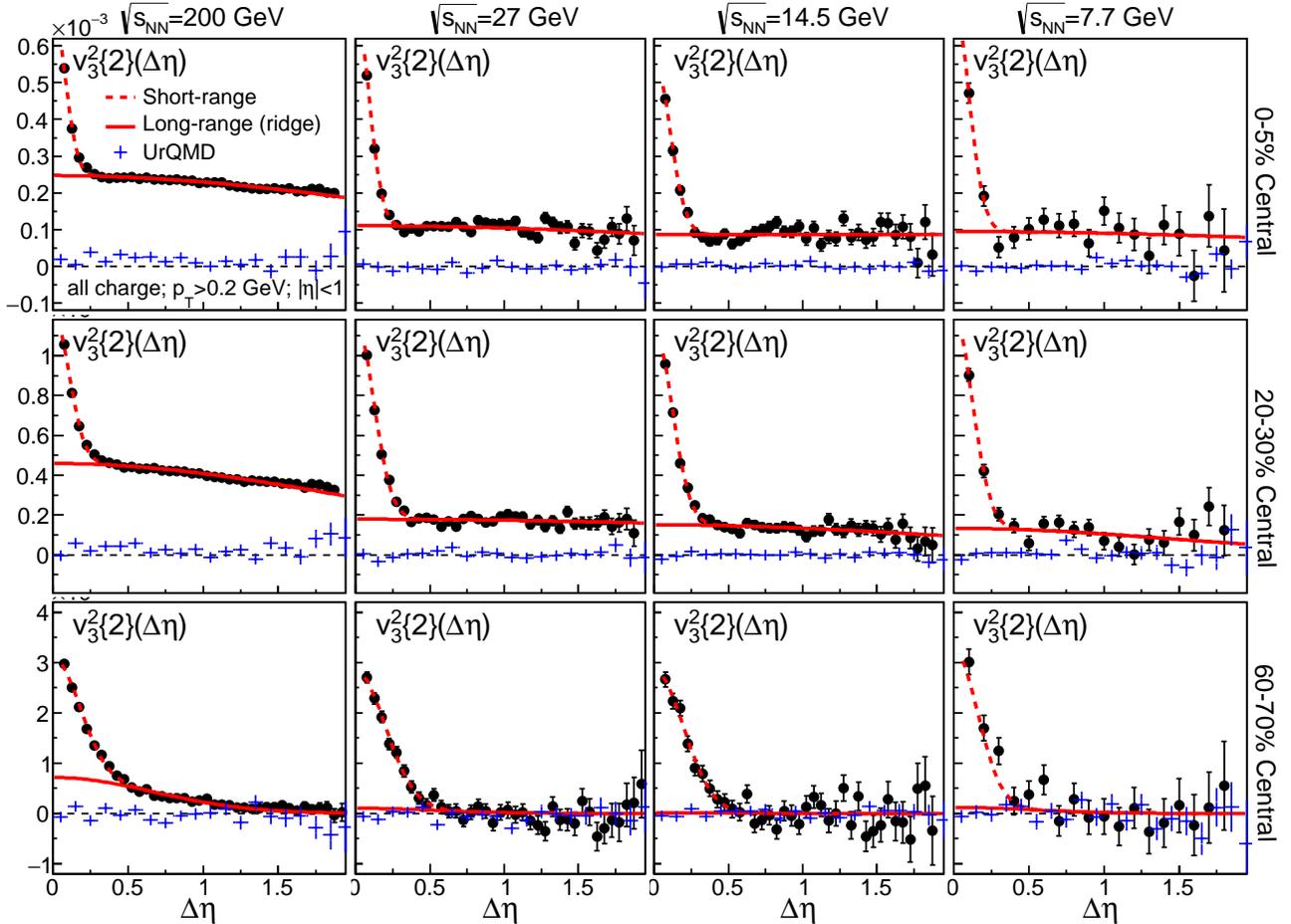}
\caption{\label{fig:deta} (Color online) Representative results on
  {\vthree} from Au+Au collisions as a function of $\Delta\eta$ for
  charged hadrons with $p_T>0.2$ GeV/c and $|\eta|<1$. The columns
  (from left to right) show data from {\snn} = 200, 27, 14.5, and 7.7
  GeV while the rows (from top to bottom) show data from 0\%-5\%,
  20\%-30\%, and 60\%-70\% centrality intervals. The error bars show
  statistical uncertainties only. The fitted curves are described in
  the text. UrQMD~\cite{Bass:1998ca} results are also shown. }
\end{figure*}

In Fig.~\ref{fig:deta}, we show examples of the third harmonic of the
two-particle azimuthal correlation functions as a function of
$\Delta\eta$ for three centrality intervals (0\%--5\%, 20\%--30\%, and
60\%--70\%) and four energies ({\snn} = 200, 27, 14.5, and 7.7
GeV). The harmonic $v_3^2\{2\}$ exhibits a narrow peak in $\Delta\eta$
centered at zero. For the more central collisions, non-zero
$v_3^2\{2\}$ persist out to large values of $\Delta\eta$.
The non-zero values of {\vthree} at larger $\Delta\eta$ are the result
of a long-range correlation phenomena called the ridge which was first
discovered in 200 GeV collisions at RHIC~\cite{ridgedata}.  In central
collisions, we observe that this long-range structure persists down to
7.7 GeV, the lowest beam energies measured at RHIC.  In peripheral
collisions, quantum interference effects grow broader owing to the
inverse relationship between the size of the system and the width of
the induced correlations. In peripheral collisions at 200 GeV, we
observe an additional residual $v_3^2\{2\}$ that, while not as wide as the ridge
in central collisions, is still too wide to be attributed to quantum
interference. At the lower beam energies however, the only
$v_3^2\{2\}$ signal present is at small $\Delta\eta$
and the ridge-like structure is absent. These data indicate that for
more central collisions, the ridge first seen at 200 GeV persists down
to the much lower energies probed in the RHIC beam energy scan. In the
peripheral collisions however, the ridge is absent at the lowest
energies. The figure also shows calculations from
UrQMD~\cite{Bass:1998ca}, a hadronic cascade model with no QGP
phase. Although UrQMD produces a significant $v_2$ in quantitative
agreement with measurements at {\snn}$<20$ GeV~\cite{Petersen:2006vm},
the model produces no appreciable $v_3$. The long-range correlations
seen in Fig.~\ref{fig:deta} are only consistent with this hadronic
model for peripheral collisions at the lower energies.

Short range correlations can arise from several sources including the
fragmentation of hard or semi-hard scattered partons
(jets)~\cite{Adams:2006yt}, from resonances, from quantum interference
(HBT)~\cite{Lisa:2005dd}, and from coulomb interference.  In central
collisions, a narrow peak arising primarily from HBT is present that
is easy to isolate from other correlations. In order to study the
remaining, longer-range correlations of interest in this letter, we
simultaneously fit that short range correlation with a narrow Gaussian
peak and the remaining correlations with a wider Gaussian with a
constant offset.  The fitting functions are shown in the figures where
the solid curves represent the correlations of interest and the dashed
curves represent the totals. We then extract {\vthree} averaged over
$\Delta\eta$ by excluding the contribution parameterized by the narrow
short-range Gaussian and integrating over the remaining structure
within $|\Delta\eta|<2$:
\begin{equation}
\langle v_3^2\{2\}\rangle = \frac{\int
(dN/d\Delta\eta)(v_3^2\{2\}(\Delta\eta)-\delta)d\Delta\eta}{\int
(dN/d\Delta\eta)d\Delta\eta}
\end{equation}
where $dN/d\Delta\eta$ is the number of pairs in each $\Delta\eta$ bin
(which decreases approximately linearly with $\Delta\eta$ to zero at
the edge of the acceptance) and $\delta$ is the contribution from the
narrow Gaussian. This quantity is extracted using the same procedure
for different centralities and different beam energies. Our analysis
does not attempt to isolate correlations attributed to flow from those
attributed to other sources like jets and resonance decays (flow vs.
non-flow)~\cite{Borghini:2001vi,Abdelwahab:2014sge}. Those non-flow
correlations typically decrease with increasing multiplicity, and thus are
not the dominating contribution in central collisions. This is
especially so for the cases where $v_3^2\{2\}$ is present in central
collisions but absent in peripheral.


\begin{figure}[htb]
\includegraphics[width=0.5\textwidth]{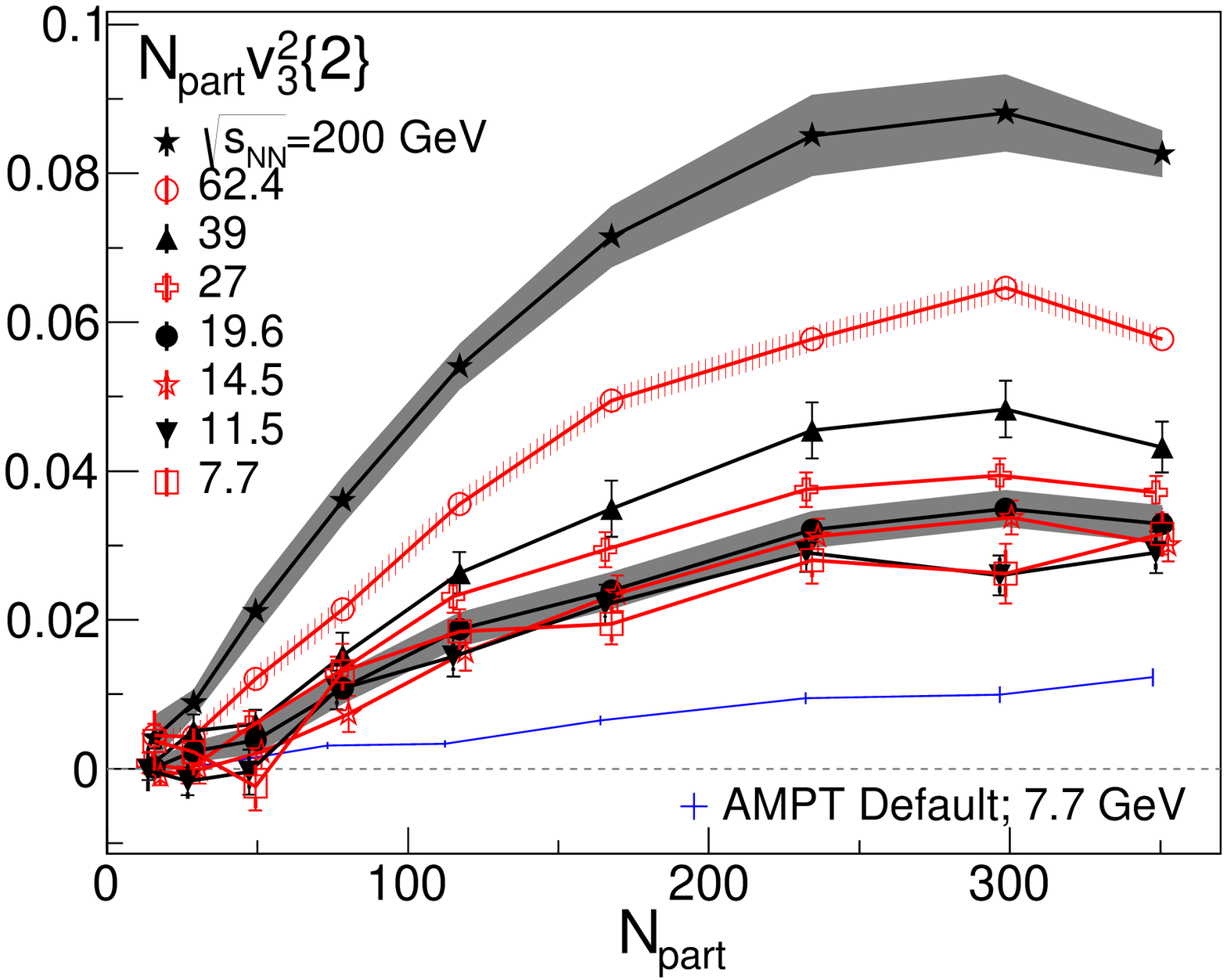}
\caption{\label{fig:cent}The {\vthree} results from Au+Au collisions
  integrated over all $\Delta\eta$ and multiplied by {\npart}. Statistical
  errors are typically smaller than the symbol size. Systematic errors
  are shown either as a shaded band or as thin vertical error bars
  with caps. The {\vthree} from a non-QGP based model, AMPT (Default),
  is also shown for {\snn}=7.7 GeV for
  comparison~\cite{Solanki:2012ne}.}
\end{figure}

In Fig.~\ref{fig:cent}, we present {\vthree} for charged hadrons
integrated over $p_T$$>$$0.2$ GeV/c and $|\eta|$$<$$1$, multiplied by
{\npart} and plotted vs. {\npart}. The figure shows data for eight
{\snn} values ranging from 7.7 to 200 GeV and for nine different
centrality intervals corresponding to 0-5, 5-10, 10-20, 20-30, 30-40,
40-50, 60-70, and 70-80\% most central. The corresponding average
{\npart} values are estimated to be 350.6, 298.6, 234.3, 167.6, 117.1,
78.3, 49.3, 28.2 and 15.7~\cite{Abelev:2008ab}. {\npart} only weakly
depends on energy and we use the same {\npart} values for all energies
even though centrality resolution changes with {\snn}.  We plot
{\npvthree} to cancel the approximate 1/{\npart} decrease one expects
for two-particle correlations or fluctuations as {\npart}
increases. If a central collision was a trivial linear superposition
of p+p collisions, then {\npvthree} would remain constant with
centrality. The data deviate drastically from the trivial expectation.
In peripheral collisions, {\npvthree} is close to zero, but then
increases with centrality until it saturates at values close to \npart
=300 before exhibiting a systematic tendency to drop slightly in the
most central bins. This drop in the most central bin is there for all
except the lowest energies where error bars become somewhat larger and
the centrality resolution becomes worse. This rise and then fall has
been traced to the non-trivial evolution of the initial geometry of
two overlapping nuclei~\cite{Sorensen:2011hm}; when the collisions are
off-axis, the effect of fluctuations in positions of nucleons in one
nucleus are enhanced when they collide with the center of the other
nucleus (increasing {\vthree}). This effect subsides when the two
nuclei collide nearly head-on. The increase of {\npvthree} is
exhibited at all energies including 7.7 GeV. Several models suggest
that the absence of a QGP should be accompanied by a significant
decrease in {\vthree}~\cite{Auvinen:2013sba,Solanki:2012ne}, but we do
not see that decrease. We include a comparison of the AMPT (Default)
hadronic model to the 7.7 GeV data~\cite{Solanki:2012ne}. The non-QGP
model predicts a smaller {\vthree} value than the data, suggesting
that a QGP phase may exist in more central collisions at energies as
low as 7.7 GeV.

Systematic errors on the integrated {\vthree} are studied by analyzing
data from different years or from different periods of the run, by
selecting events that collided at different z-vertex positions, by
varying the efficiency correction within uncertainties, and by varying
the selection criteria on tracks. A systematic uncertainty is also
assigned based on the fitting and subtraction of the short range
correlations (we assume a 10\% uncertainty on the subtraction) and on
residual acceptance corrections (10\% of
$\langle\cos3\phi\rangle^2+\langle\sin3\phi\rangle^2$). These errors
are all added in quadrature for the final error estimate.

\begin{figure}[htb]
\includegraphics[width=0.5\textwidth]{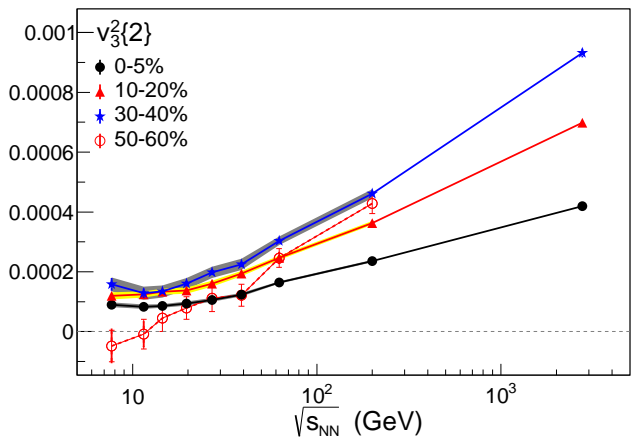}
\caption{\label{fig:edep} The {\snn} dependence of {\vthree} for four
  representative centrality intervals. All data are Au+Au except for
  the 2.76 TeV data points from the ALICE collaboration~\cite{earlyv3}
  which are Pb+Pb. ALICE data are not available for the 50\%-60\%
  centrality interval.}
\end{figure}

In Fig.~\ref{fig:edep}, we re-plot the data from Fig.~\ref{fig:cent}
for several centralities as a function of {\snn}. Data from 2.76 TeV
Pb+Pb collisions are also included~\cite{earlyv3}. At 200 GeV, the
50\%-60\% central data are similar to the 30\%-40\% data. As the
collision energy decreases however, values in the peripheral 50\%-60\%
centrality data group drop well below the 30\%-40\% central data and
become consistent with zero for 7.7 and 11.5 GeV collisions. This
shows again, that peripheral collisions at lower energies seem to fail
to convert geometry fluctuations into a ridge-like correlation. This
idea is consistent with the absence of a low viscosity QGP phase in
low energy peripheral collisions~\cite{Auvinen:2013sba}. For more
central collisions however, {\vthree} is finite even at the lowest
energies and changes very little from 7.7 GeV to 19.6 GeV. Above that,
it begins to increase more quickly and roughly linearly with
$\log(\sqrt{s_{_{\rm NN}}})$. This trend continues up to 2.76 TeV
where for corresponding centrality intervals, the {\vthree} values are
roughly twice as large as those at 200 GeV. Given that the dominant
trend at the higher energies is for {\vthree} to increase with
$\log(\sqrt{s_{_{\rm NN}}})$, it is notable that {\vthree} is
approximately constant for the lower energies. 

One would expect, independent of what energy range is considered, that
higher energy collisions producing more particles should be more
effective at converting initial state geometry fluctuations into
{\vthree}. Deviations from that expectation could indicate
interesting physics like a softening of the
equation-of-state~\cite{Sorge:1998mk}. We investigated these
expectations at the lower {\snn} by scaling {\vthree} by the
mid-rapidity, charged-particle multiplicity, pseudo-rapidity
  density per participant-pair
$n_{\mathrm{ch,PP}}=\frac{2}{N_{\mathrm{part}}}dN_{\rm ch}/d\eta$.

\begin{figure}[htb]
\includegraphics[width=0.5\textwidth]{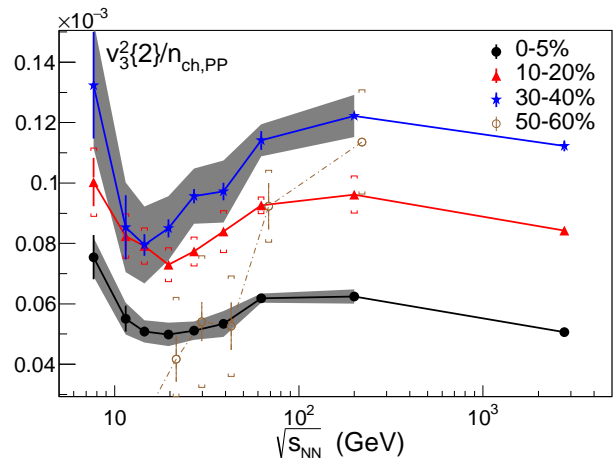}
\caption{\label{fig:scaled}{\vthree} divided by the mid-rapidity,
  charged particle multiplicity, pseudo-rapidity density per
  participant pair in Au+Au and Pb+Pb (2.76 TeV) collisions. Data in
  the centrality range from 0-50\% exhibit a local minimum near 20 GeV
  while the more peripheral events do not.}
\end{figure}

We parameterize the {\snn} dependence of the existing data on
$n_{\mathrm{ch,PP}}$ for central Au+Au or Pb+Pb collisions~\cite{mult}
by
\begin{equation}
n_{\mathrm{ch,PP}} = 
 \left\{ \begin{array}{lr}
  0.77(\sqrt{s_{_{\rm NN}}})^{0.30} & \mbox{$\sqrt{s_{_{\rm NN}}}>16.0$ GeV } \\
  0.78\log(\sqrt{s_{_{\rm NN}}}) - 0.4 & \mbox{otherwise}
\end{array} \right.
\end{equation}
In Fig.~\ref{fig:scaled}, we show $v_{3}^{2}\{2\}/n_{\mathrm{ch,PP}}$
for four centrality intervals. The more central data exhibit a local
minimum in the {\snn} range around 15-20 GeV which is absent for
peripheral collisions. Variations of
$v_{3}^{2}\{2\}/n_{\mathrm{ch,PP}}$ with different parameterizations
of $n_{\mathrm{ch,PP}}$ are typically on the order of a few
percent. The trends in $n_{\mathrm{ch,PP}}$ also have a change in
  behavior in the same energy range where the dip appears in
  Fig.~\ref{fig:scaled}, but the apparent minima in the figure do not
depend on the details of the parameterization of
$n_{\mathrm{ch,PP}}$; the local minima remain even if scaling by
  $log(\sqrt{s_{_{NN}}})$.  The minima are an inevitable consequence
of the near independence of {\vthree} with respect to {\snn} for
{\snn}$<20$ GeV while simultaneously, the multiplicity is
monotonically increasing. If the otherwise general increase of
{\vthree} is driven by ever increasing pressure gradients in ever
denser systems at higher energies, then the local minimum in
$v_{3}^{2}\{2\}/n_{\mathrm{ch,PP}}$ could be an indication of an
anomalously low pressure inside the matter created in collisions with
energies near 15-20 GeV. We note that the
  minima in Fig.~\ref{fig:scaled} could depend on the specific
  scaling scheme and more rigorous theoretical modelling is needed to
  connect this measurement to the initial density and
  flow dynamics.  In addition, the interpretation of data in this
energy range is complicated by changes in the baryon to meson
ratio~\cite{BraunMunzinger:2001as}, a relatively faster increase of
$\mu_B$ driven by baryon stopping~\cite{Arsene:2009aa}, possible
  changes in the sources and magnitude of
  non-flow~\cite{Abdelwahab:2014sge}, and the longer crossing times
for nuclei at lower energies~\cite{Auvinen:2013sba}. The existence of
the minimum in $v_{3}^{2}\{2\}/n_{\mathrm{ch,PP}}$ and other
provocative trends in data collected around these energies including
the minimum in the slope of the net proton
$v_1$~\cite{Adamczyk:2014ipa} is interesting and provides ample
motivation for further investigation~\cite{Heinz:2015tua}.

In summary, we presented measurements of the {\snn} dependence of
{\vthree} in Au+Au collisions for {\snn} energies ranging from 7.7 to
200 GeV. The conversion of density fluctuations in the initial-state
have previously been found to provide a simple explanation for
{\vthree} and the corresponding ridge correlations. Model calculations
have shown that while $v_{2}$ can also be established over a longer
period in a higher viscosity hadronic phase, {\vthree} is particularly
sensitive to the presence of a low viscosity plasma phase in the
evolution of the collision. By studying the $\Delta\eta$ dependence of
{\vthree}, we find that for sufficiently central collisions ({\npart}
$>50$), the ridge and {\vthree} persist down to the lowest energies
studied. For more peripheral collisions however, the ridge correlation
appears to be absent at low energies for \npart $<50$, in
  agreement with certain non-QGP models. When comparing {\vthree} at
RHIC and the LHC, the much larger multiplicities at the LHC lead to a
much larger {\vthree}. When divided by multiplicity, $v_3^2\{2\}$
shows a local minimum in the region near 15-20 GeV. This feature has
not been shown in any known models of heavy ion collisions and could
indicate an interesting trend in the pressure developed inside the
system.


We thank the RHIC Operations Group and RCF at BNL, the NERSC Center at
LBNL, the KISTI Center in Korea, and the Open Science Grid consortium
for providing resources and support. This work was supported in part
by the Office of Nuclear Physics within the U.S. DOE Office of
Science, the U.S. NSF, the Ministry of Education and Science of the
Russian Federation, NSFC, CAS, MoST and MoE of China, the National
Research Foundation of Korea, NCKU (Taiwan), GA and MSMT of the Czech
Republic, FIAS of Germany, DAE, DST, and UGC of India, the National
Science Centre of Poland, National Research Foundation, the Ministry
of Science, Education and Sports of the Republic of Croatia, and
RosAtom of Russia.


\end{document}